\documentclass[useAMS,usenatbib]{mn2e}
\usepackage{hyperref}
\usepackage{graphicx}
 
\begin{document}

\title[ Cold galaxies]{Cold galaxies
\thanks{{\it Herschel} is an ESA space observatory with science
instruments provided by European-led Principal Investigator
consortia and with important participation from NASA.}
}
\author[Rowan-Robinson M. and Clements D.L.]{Michael Rowan-Robinson$^{1}$, David L. Clements$^{1}$\\
\thanks{E-mail: \texttt{mrr@imperial.ac.uk}},
${1}$Astrophysics Group, Imperial College London, Blackett Laboratory, Prince Consort Road, London SW7 2AZ, UK\\
}
\maketitle
\begin{abstract}
We use 350 $\mu$m angular diameter estimates from {\it Planck} to test the idea that some galaxies
contain exceptionally cold (10-13 K) dust, since colder dust implies a lower surface brightness
radiation field illuminating the dust, and hence a greater physical extent for a given
luminosity.  The galaxies identified from their spectral energy
distributions as containing cold dust do indeed show the expected larger 350 $\mu$m diameters.
For a few cold dust galaxies where {\it Herschel} data are available
we are able to use submillimetre maps or surface brightness profiles to
locate the cold dust, which as expected generally lies outside the optical galaxy.

\end{abstract}


\section{Introduction}
One of the surprising results from submillimetre surveys with  {\it Herschel} and {\it Planck} has
been the discovery that some local
(z$<$0.1) quiescent galaxies show spectral energy distributions (SEDs) characteristic of
cold (T=10-13 K) dust (Rowan-Robinson et al 2011, 2014, Ade et al 2011, Wang and Rowan-Robinson 2014). 
Two examples are illustrated in Fig 1.

Cold dust in submillimetre galaxies has also been investigated by Galametz et al (2012), Smith et al (2012) and
Symeonidis et al (2013) using SED fitting, by Bourne et al (2013) using CO and by Ibar et al (2013) using CII.
Bendo et al (2014) have used surface brightness ratios at 250, 350 and 500 $\mu$m to explore the relative
contribution of newly formed and old stars to dust heating.

\begin{figure*}
\includegraphics[width=7cm]{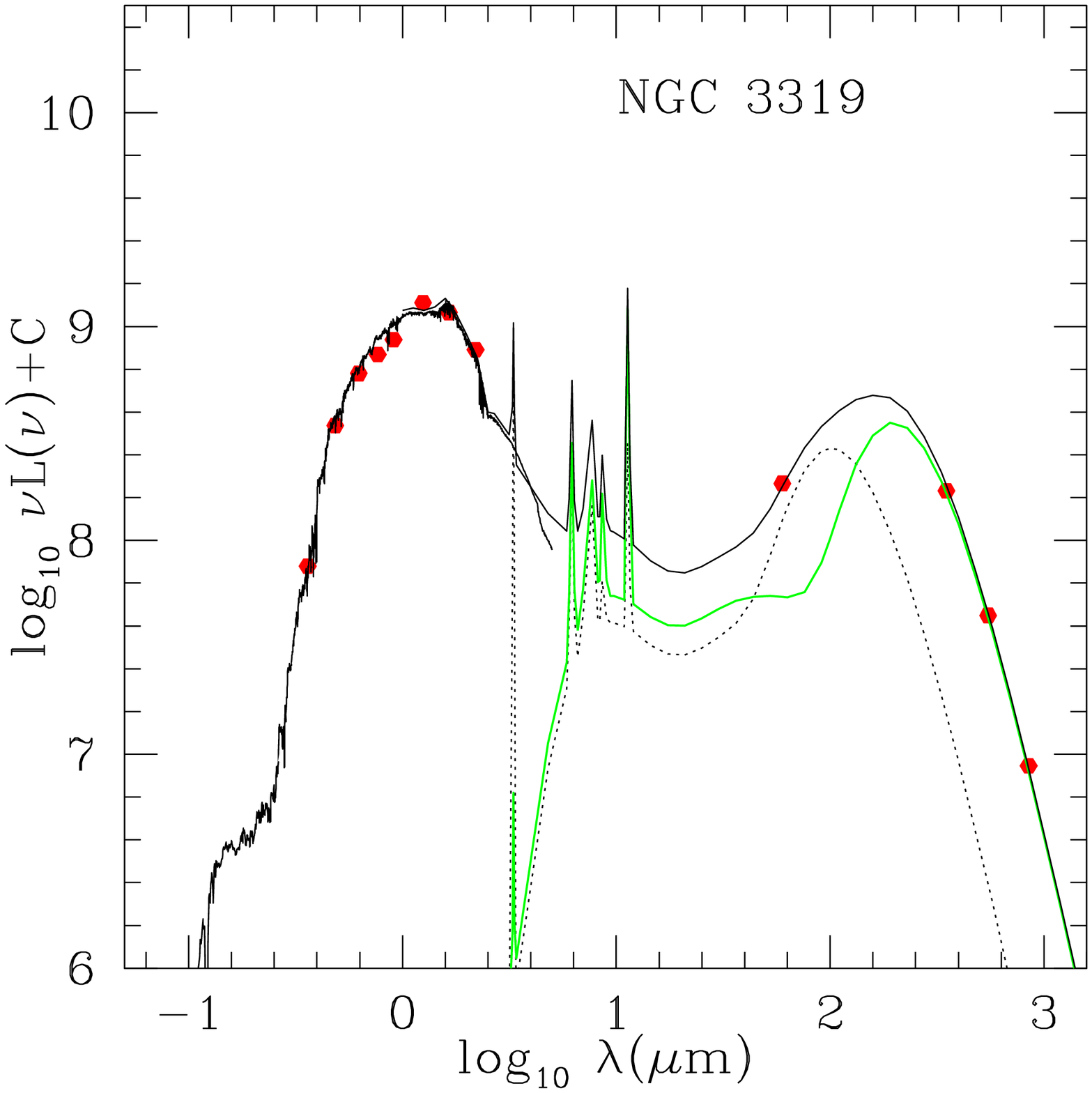}
\includegraphics[width=7cm]{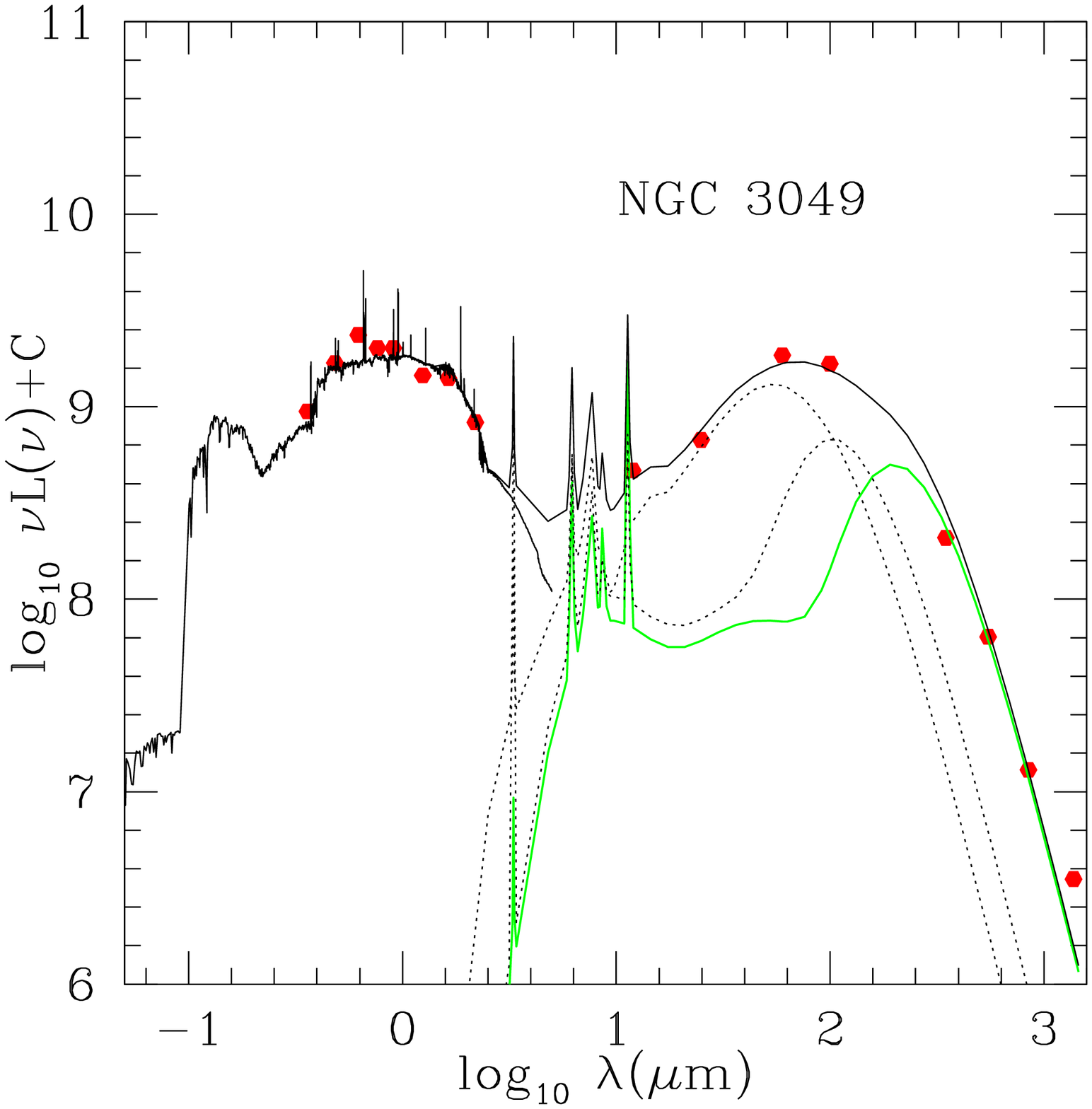}
\caption{
SEDs of NGC 3319 and NGC 3049, showing need for cold dust component (green).
}
\end{figure*}

To test the idea that exceptionally cold dust is present in some galaxies we compare {\it Planck} 350 $\mu$m diameters with diameters estimated from
radiative transfer models for the far infrared and submillimetre emission.  We also look at submillimetre
maps for selected galaxies derived from {\it Herschel} data.  Broadly speaking we expect to find
that galaxies with colder dust will be more extended spatially at submillimetre wavelengths than
normal galaxies of the same luminosity.

\begin{figure*}
\includegraphics[width=14cm]{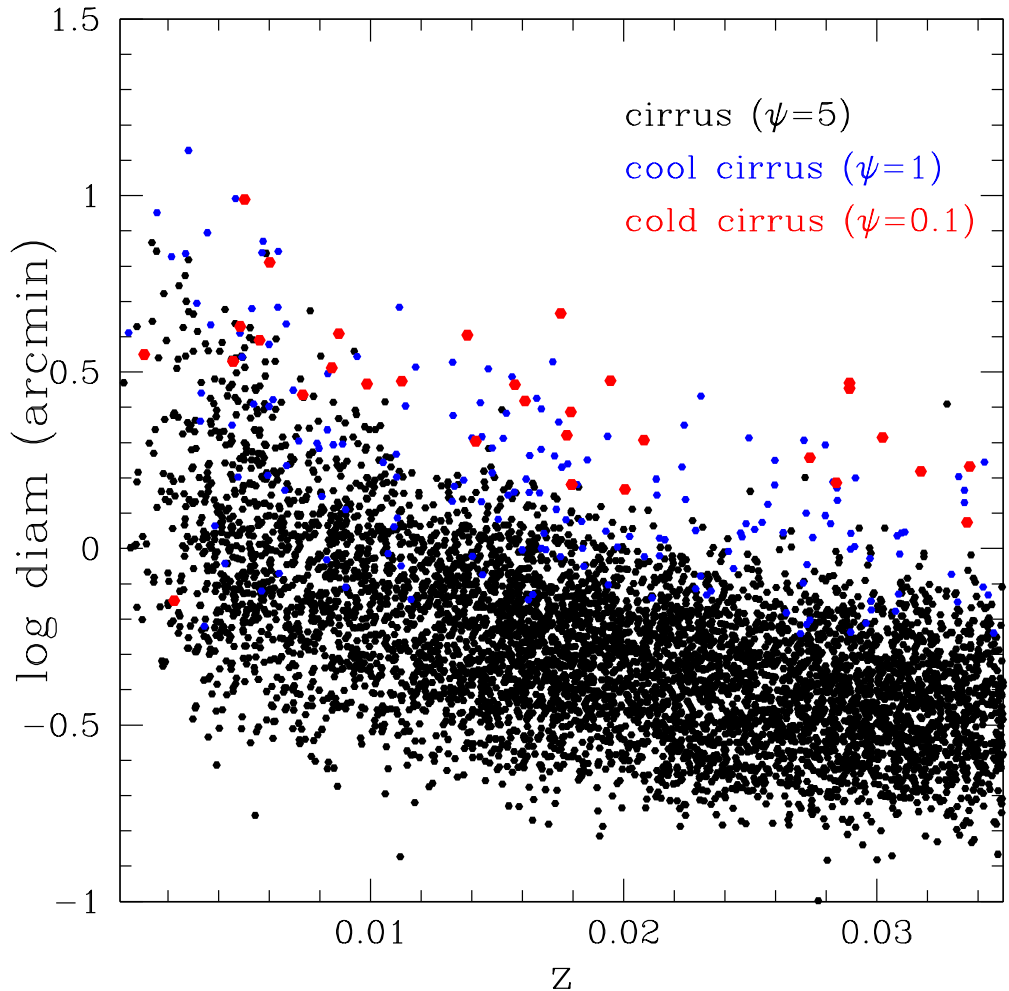}
\caption{Predicted submillimetre diameter versus redshift for cirrus ($\psi$=5) galaxies (black),
cool galaxies ($\psi$=1, blue) and cold galaxies ($\psi$=0.1, red).
}
\end{figure*}

The structure of this paper is as follows: in section 2 we estimate the submillimetre diameters
for galaxies with cool and cold dust and compare these with predictions for normal cirrus
galaxies, in section 3 we compare these predictions with the 350 $\mu$m diameters observed
by {\it Planck} and in section 4 we discuss the few cases where we have {\it Herschel} maps for cold
dust galaxies.  Section 5 gives our discussion and conclusions.

\section{Predicted submillimetre diameters}

The temperature of optically thin interstellar dust ('cirrus') is determined by the
intensity of the interstellar radiation field, which can be characterised by
the ratio of the intensity of the radiation field to the local Solar Neighbourhood interstellar 
radiation field, $\psi$ (Rowan-Robinson et al 1992, hereafter RR92).
The standard cirrus template of Rowan-Robinson et al (2005, 2008, 2013) corresponds to $\psi$ = 5, and this
is the value used by RR92 to fit the central regions of our Galaxy.  
$\psi$ = 1 corresponds to the interstellar radiation field in the vicinity of the Sun ('cool' dust).
Rowan-Robinson et al (2010, 2014) found that some {\it Herschel} galaxies need a much lower intensity radiation
field than this, with $\psi$ = 0.1 ('cold' dust).   A similar result was found for some {\it Planck}
galaxies by Ade et al (2011).  The corresponding grain temperatures in the dust
model of RR92 are given in Table 1 of Rowan-Robinson et al (2010).  For the
three values of $\psi$ = 5, 1, 0.1, the ranges of dust grain temperatures for the different grain types 
are 20-24 K, 14.5-19.7 K and 9.8-13.4 K respectively.  Full details of the templates used are given 
via a readme page
\footnote{\url{http://astro.ic.ac.uk/public/mrr/swirephotzcat/templates/readme}}.

One of the ways to test these ideas is to measure the submillimetre diameter. Since $\psi$ = 1
corresponds to the intensity in the solar neighbourhood, we can estimate the submillimetre diameter of
galaxies by
\newline

	$\theta \sim 2 R_0 \psi^{-0.5} (z c\tau_0)^{-1} (L_{ir}/L_{MW})^{-0.5}$   (1)
\newline

where $R_0$ = 8.5 kpc, the distance of the Sun from the Galactic Centre, and 
$L_{MW} = 2.4 . 10^{10} L_{\odot}$.
The definition of $\psi$ in the Solar Neighbourhood is in terms of a mean surface brightness and so we can
expect that $\theta$ defined by this equation is essentially a full-width to
half-power when observing external galaxies.

Figure 2 shows a plot of the predicted angular size, $\theta$, versus redshift for
galaxies in the RIFSCz IRAS galaxy redshift catalogue (Wang and Rowan-Robinson 2014) whose far infrared and 
submillimetre emission is best fitted by standard cirrus or cool cirrus, for the
sample of {\it Planck} galaxies identified as requiring cold dust by Ade et al (2011), and for
the RIFSCz galaxies identified as needing cold dust templates by Wang and Rowan-Robinson (2014).
The prediction is that galaxies with significant cool or cold dust components should have larger
submillimetre diameters than their normal cirrus counterparts.
Table 1 summarises the properties of the cold dust galaxies from these latter two papers.

\section{Measured {\it Planck} 350 $\mu$m diameters}
We have focussed on those galaxies for which the predicted subillimetre diameter is
greater than 3 arcmin and compared these with the PCCS submillimetre diameters 
(FWHM) observed by {\it Planck} at 350 $\mu$m.
We use the Gaussian half-power width measured along the major axis.
Table 2 summarises the properties of galaxies from  the IRAS RIFSCz catalogue
with predicted submillimetre diameter greater than 3 arcmin, spectroscopic redshifts, at least
8 optical photometric bands, $\chi^2$ for the infrared template fit $<$ 5, and SEDs dominated
either by a cirrus or cool dust template. Figure 3L shows a comparison of the observed {\it Planck} diameters with
the optical diameters, taken from the 3rd Reference Catalogue of Galaxies, for the galaxies of
Tables 1 and 2.  The quoted Planck beam at 350 $\mu$m is 4.3 arcmin (FWHM), so we deconvolve
both the telescope beam and source profile can be approximated as Gaussians.  Figure 3R shows a
the observed diameters using $\theta_{deconv} = (\theta^2 - 4.3^2)^{-1/2}$, assuming that
both the telescope beam and the source profile can be approximated as Gaussians (adaquate
for our purposes).  Values of $\theta_{deconv} < 4$ arcmin will be subject to considerable uncertainty.
Figure 3R shows a comparison of the deconvolved
{\it Planck} diameters with the predicted values
from eqn (1).  All three types are consistent with a simple linear relation, apart
from a cluster of galaxies with anomalously high {\it Planck} diameters, which we suggest are
affected by Galactic cirrus (most have high cirrus flags in the {\it Planck} PCCS, eg NGC 1024, 3573, 4650, and IC 4831 and 5078).
We have fitted straight lines through the origin to these three distributions and find
slopes 1.38$\pm$0.13 for 40 normal cirrus galaxies, 1.16$\pm$0.09 for 64 cool galaxies
(excluding 8 probable cirrus sources) and 1.32$\pm$0.21 for 8 cold galaxies.  Thus all 
three populations show statistically significant correlations between the observed and
predicted diameters, the slopes are consistent with being the same in each case and
are also not inconsistent with the true slope being unity.

There are also two outliers with rather large predicted diameters, NGC 6744 and M 104.
Figures 4L and 5 show the SEDs for these two outliers.  For NGC6744, once the
IRAS fluxes from the IRAS Large Galaxy Catalog (Rice et al 1988) have been used, the
SED is in fact dominated by a standard cirrus template, so it should move to the left
in Fig 3R.
For M104, Bendo et al (2006) have suggested that the AGN at the centre of the galaxy contributes to the
submillimetre emission.
Figure 4R shows {\it Herschel} 250 $\mu$m contours superposed on an i-band image for M104 ('The Sombrero').  The
250 $\mu$m emission arises from a ring of dust outside the main distribution of starlight.
The predicted value of $\theta$ may also be an overestimate because in this edge-on system
the starlight illuminating the dust ring has been significantly attenuated by dust in
the disk of the galaxy, so again this object should move to the left in Fig 3R.

\begin{figure*}
\includegraphics[width=7cm]{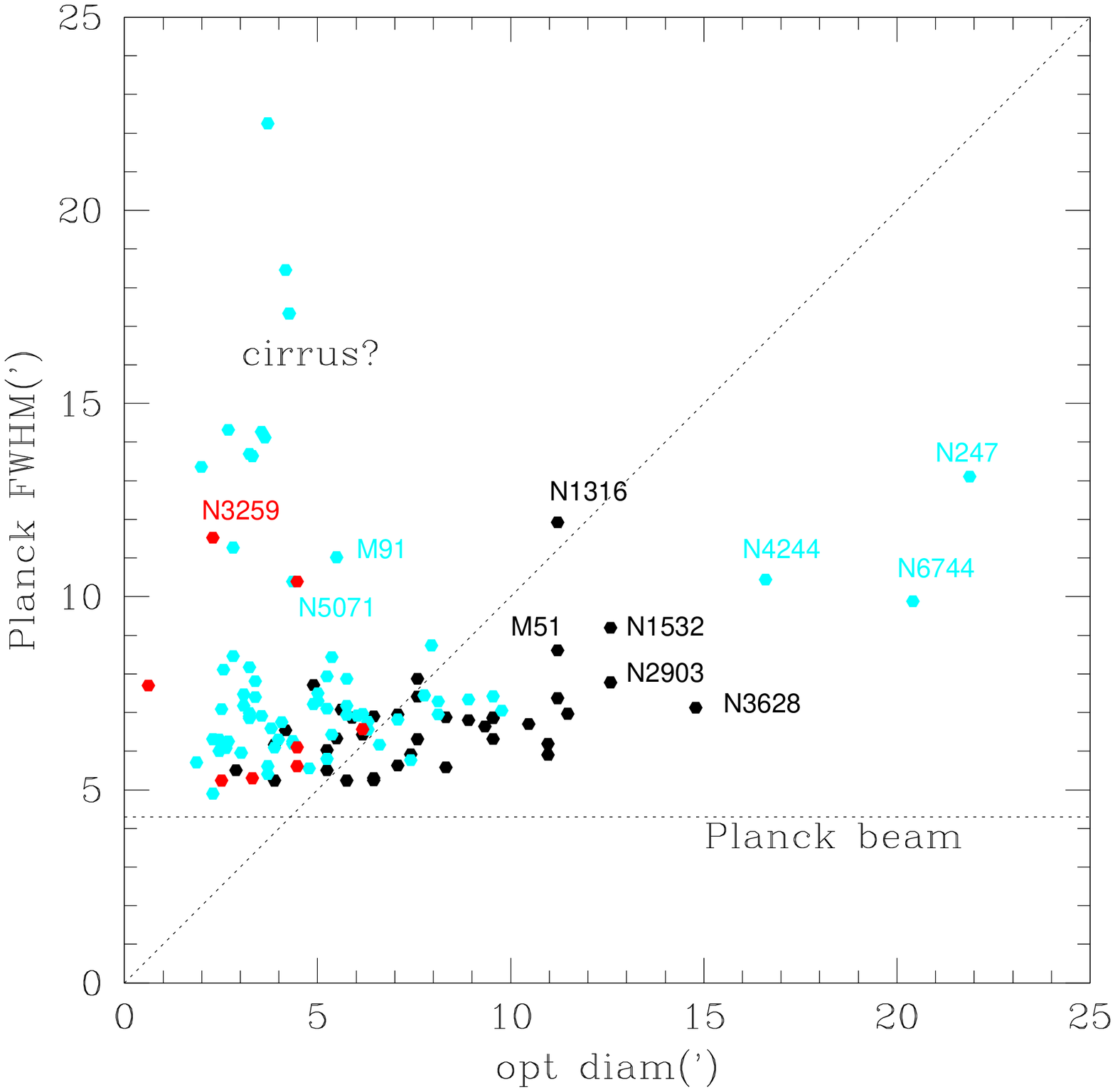}
\includegraphics[width=7cm]{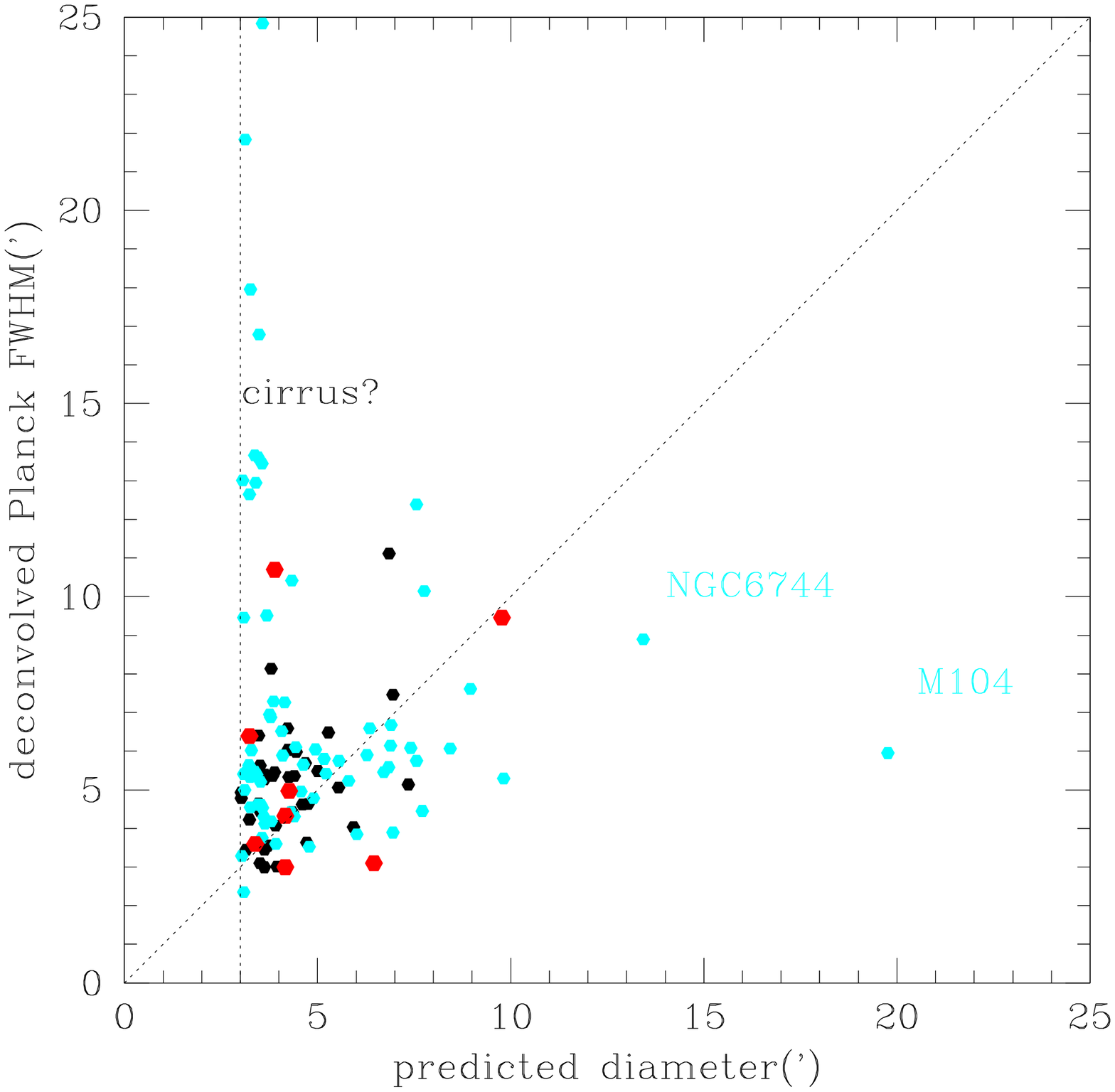}
\caption{
L: Observed {\it Planck} 857 GHz FWHM versus optical diameter (arcmins) for standard cirrus
galaxies ($\psi$=5, black), cool cirrus galaxies ($\psi$=1, cyan) and cold cirrus galaxies ($\psi$=0.1, red).
The dotted line denotes the Planck 857 GHz beam (FWHM).
R: Deconvolved {\it Planck} 857 GHz FWHM versus predicted submillimetre diameter (arcmins) for standard cirrus
galaxies ($\psi$=5, black), cool cirrus galaxies ($\psi$=1, cyan) and cold cirrus galaxies ($\psi$=0.1, red).}
Here the dotted line indicates the selection cutoff $\theta$=3.0.
\end{figure*}

\begin{figure*}
\includegraphics[width=6cm]{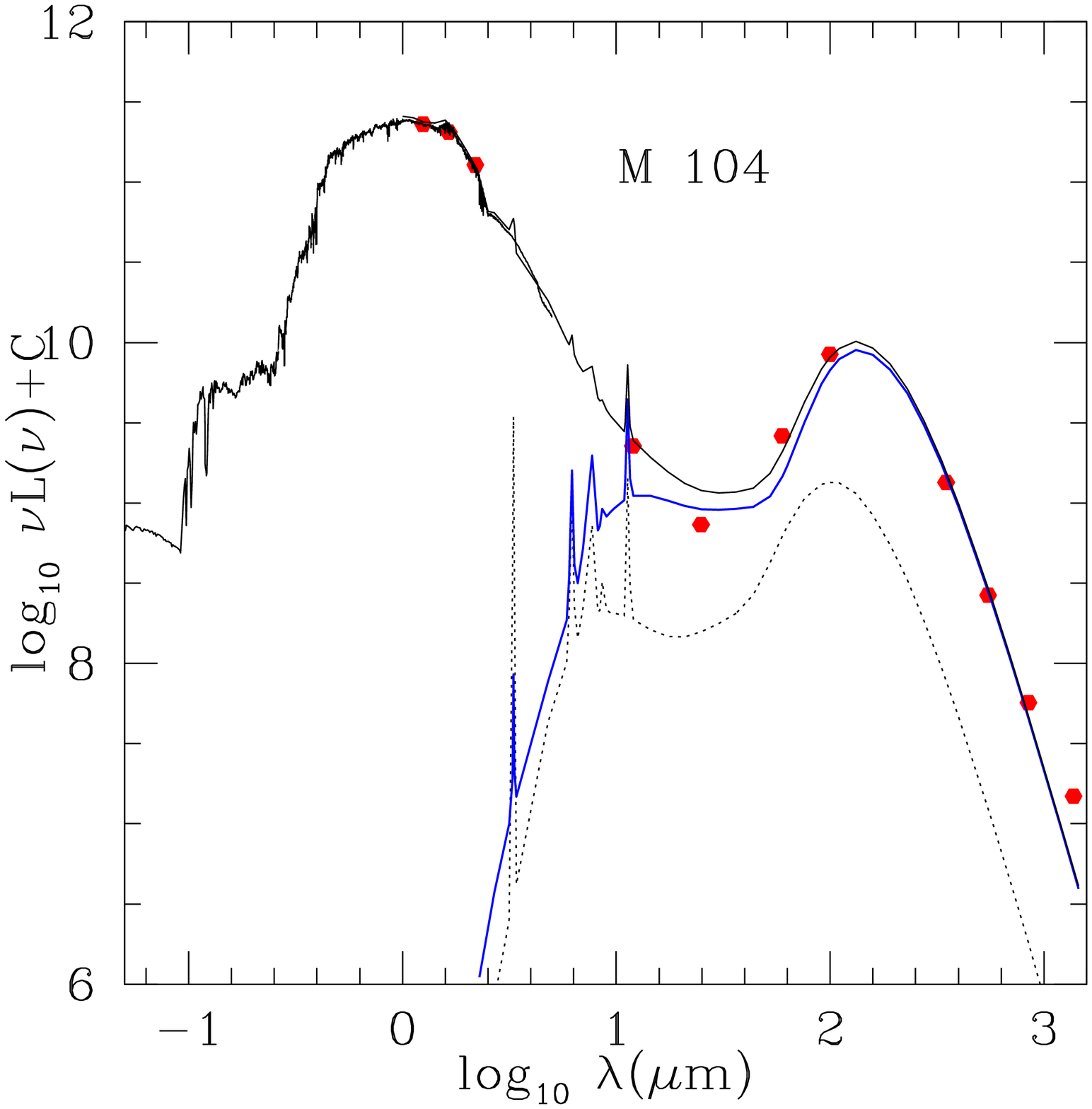}
\includegraphics[width=8cm]{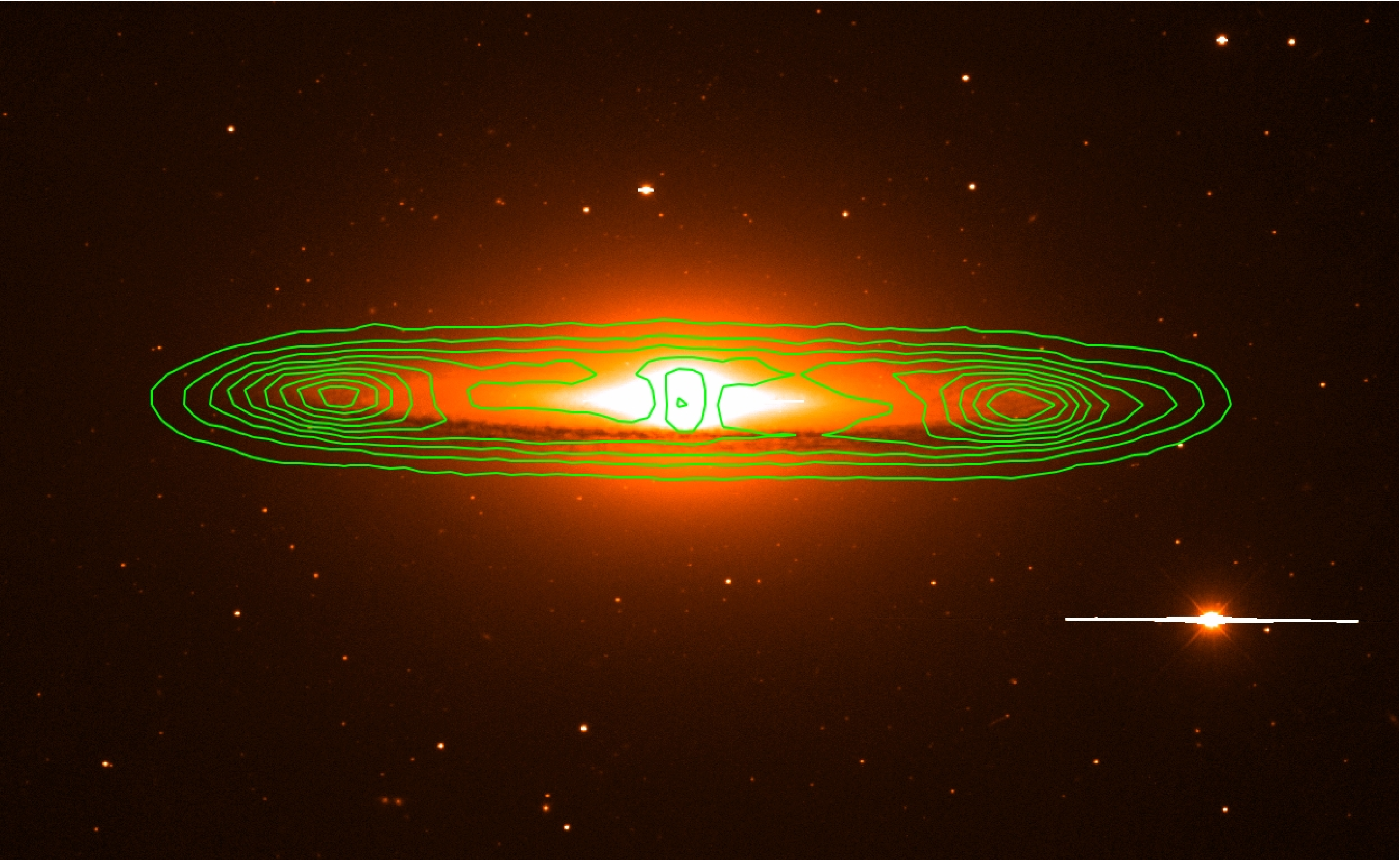}
\caption{
L: SED of M104, an outlier in Fig 3R.
R: 250 $\mu$m contour map of M104 superposed on i-band image.  Much of the 250 $\mu$m emission
arises from a ring of dust outside the main optical image.
}
\end{figure*}

\begin{figure}
\includegraphics[width=7cm]{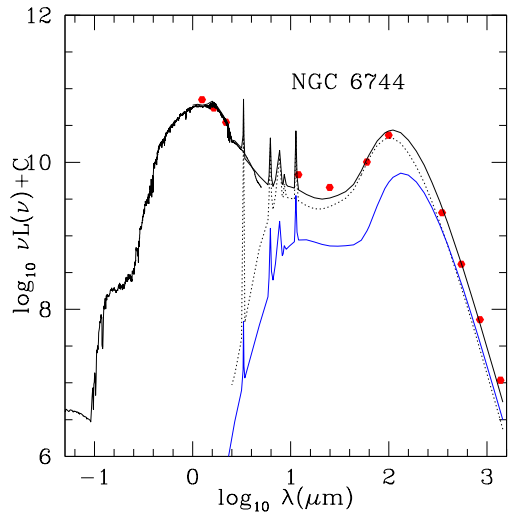}
\caption{
SED of  NGC 6744, outlier in Fig 3R.
}
\end{figure}

\begin{table*}
\caption{Galaxies requiring cold dust components}
\begin{tabular}{llllllll}
&&&&&&&\\
IRAS name & z & $L_{opt}$ & $log_{10}\theta_{pred}$ & $\theta_{Planck}$ & $log_{10}\theta_{opt}$ & ncirr & name\\
(a) {\it Planck} selected &&&&&&&\\
F10361+4155 & 0.002228 & 7.50 & 0.63 & 6.57 & 0.79 & 1 & NGC3319 \\
F14302+1006 & 0.00457 &  9.48 & 0.53 & 5.61 & 0.65 & 1 & NGC5669 \\
F09521+0930 & 0.00485 &  9.73 & 0.62 & 5.24 & 0.40 & 1 & NGC3049 \\
F10290+6517 & 0.00562 &  9.78 & 0.59 & 11.53 & 0.36 & 3 & NGC3259 \\
F14004+5603 & 0.00601 & 10.28 & 0.81 & 5.3 & 0.52 & 1 & NGC5443 \\
F15122+5841 & 0.00847 &  9.98 & 0.51 & 7.7 & -0.21 & 3 & MrK0847 \\
F15248+4044 & 0.00874 & 10.20 & 0.62 & 6.1 & 0.65 & 1 & UGC09858 \\
F14366+0534 & 0.00502 & 10.48 & 0.99 & 10.39 & 0.65 & 3 & NGC5701 \\
F13405+6101 & 0.00732 &  9.70 & 0.43 & 6.71 & 0.23 & 1 & UGC08684 \\
F08233+2303 & 0.01794 &  9.97 & -0.03 & & & & KUG0823+230B \\
F15243+5237 & 0.01948 & 10.63 & 0.36 & & & & UGC09853 \\
F15495+5545 & 0.03974 & 10.90 & -0.54 & & & & SBS1549+557 \\
F04257-4913 & 0.05828 & 11.85 & -0.10 & & & & ESO202-IG021 \\
&&&&&&&\\
(b) RIFSCz selected &&&&&&&\\
F12234+3348 &   0.001060 &  8.25 &  0.550 & & 1.12 & & NGC4395 \\  
F23461+0353 &   0.009860 & 10.02 &  0.466 & & 0.40 & & NGC7757 \\
F11555+2809 &   0.011230 & 10.15 &  0.475 & & 0.26 & & NGC4004 \\ 
F02533+0029 &   0.013820 & 10.59 &  0.604 & & 0.15 & & UGC02403 \\
F12208+0744 &   0.014160 & 10.01 &  0.304 & & 0.36 & & NGC4334 \\ 
F15097+2129 &   0.015700 & 10.42 &  0.464 & & 0.18 & & UGC09763 \\ 
F00342+2342 &   0.016110 & 10.35 &  0.418 & & 0.43 & & NGC169 (Arp282) \\  
F08070+3406 &   0.017520 & 10.92 &  0.666 & & 0.35 & & NGC2532 \\
F14280+2158 &   0.017750 & 10.24 &  0.321 & & 0.19 & & UGC09316 \\ 
F07581+3259 &   0.017910 & 10.38 &  0.387 & & -0.24 & & CGCG178-018 \\  
F08277+2046 &   0.020050 & 10.04 &  0.168 & & 0.05 & & UGC04446 \\
F09388+1138 &   0.020810 & 10.35 &  0.307 & & 0.36 & & UGC05173 \\
F13090+4658 &   0.027370 & 10.49 &  0.258 & & -0.23 & & UGC08269 \\
F14236+0528 &   0.028400 & 10.38 &  0.187 & & -0.41 & & UGC09244 \\
F23254+0830 &   0.028920 & 10.96 &  0.469 & & 0.06 & & NGC7674 \\
F08199+0427 &   0.028930 & 10.93 &  0.454 & & -0.31 & & CGCG032-009 \\  
F11078+0505 &   0.030230 & 10.69 &  0.315 & & -0.24 & & UGC06212 \\
F15426+4115 &   0.031750 & 10.54 &  0.218 & & -0.05 & & NGC5992 \\
F16269+4013 &   0.033570 & 10.30 &  0.074 & & -0.38 & & KUG1626+402 \\  
F14547+2449 &   0.033670 & 10.62 &  0.233 & & -0.18 & & UGC09618 \\
\end{tabular}
\end{table*}

\begin{figure*}
\includegraphics[width=7cm]{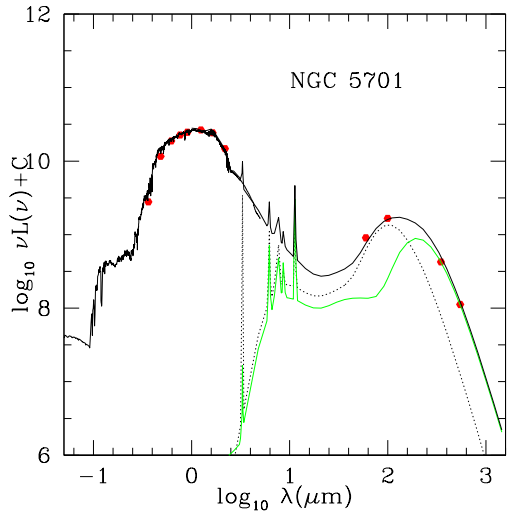}
\includegraphics[width=7cm]{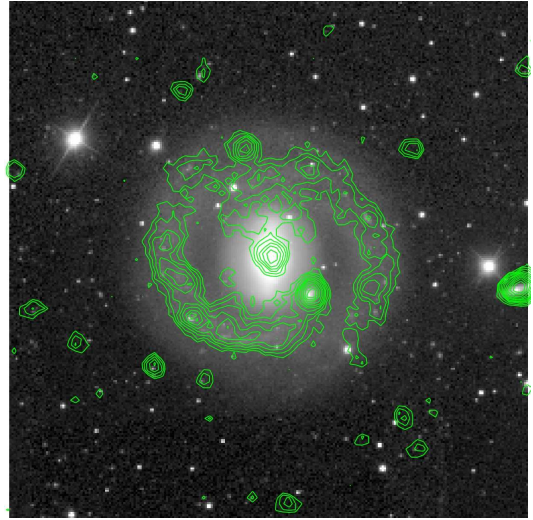}
\caption{
L: SED of NGC 5701. R: 250 $\mu$m map of NGC5701.  In this case the submillimetre
emission arises from a ring of dust lying outside the main stellar distribution. The 
peak of emission to the SW of the nucleus is a chance superposition of a 
background z=0.04 galaxy.
}
\end{figure*}

\begin{table*}
\caption{Comparison samples of large galaxies requiring cool dust or normal cirrus components}
\begin{tabular}{llllllll}
IRAS name & z & $L_{opt}$ & $log_{10}\theta_{pred}$ & $\theta_{Planck}$ & $log_{10}\theta_{opt}$ & ncirr & name\\
&&&&&&&\\
cool dust galaxies &&&&&&&\\
F00380-1408 &   0.005460 & 11.17 &  0.798 & 7.30 & 0.70 & 1 & NGC0210\\
F00446-2101 &   0.000520 &  9.29 &  0.879 & 13.11 & 1.34 & 2 & NGC0247\\
F01191+0459 &   0.007580 & 11.24 &  0.690 & 6.43 & 0.73 & 1 & NGC0488\\
F01443+3519 &   0.015590 & 11.46 &  0.487 & 13.7 & 0.51 & 4 & NGC0669\\
F02364+1037 &   0.011780 & 11.27 &  0.514 & 18.46 & 0.62 & 7 & NGC1024\\
F02403+3707 &   0.001730 &  9.60 &  0.512 & 6.85 & 0.51 & 1 & NGC1058\\
F02415-2912 &   0.004840 & 10.69 &  0.610 & 7.81 & 0.53 & 1 & NGC1079\\
F03116-0300 &   0.005710 & 10.85 &  0.618 & 8.44 & 0.73 & 1 & NGC1253\\
F03151-3245 &   0.015140 & 11.44 &  0.490 & 4.90 & 0.36 & 1 & NGC1288\\
F03174-1935 &   0.005260 & 11.07 &  0.764 & 6.77 & 0.80 & 1 & NGC1300\\
F03222-2143 &   0.005310 & 10.91 &  0.680 & 5.56 & 0.68 & 2 & NGC1325\\
F03291-3348 &   0.006350 & 11.39 &  0.842 & 5.80 & 0.72 & 1 & NGC1350\\
F03309-1349 &   0.006660 & 11.02 &  0.637 & 11.27 & 0.45 & 2 & NGC1357\\
F03367-2629 &   0.004660 & 11.42 &  0.992 & 6.82 & 0.85 & 1 & NGC1398\\
F03401-3003 &   0.005040 & 11.11 &  0.803 & 7.87 & 0.76 & 1 & NGC1425\\
F03417-3600 &   0.004630 & 10.46 &  0.514 & -9.99 & 0.47 & 5 & NGC1436\\
F03451-3351 &   0.003600 & 10.37 &  0.579 & 6.00 & 0.39 & 2 & IC1993\\
F04305-5442 &   0.003550 & 10.99 &  0.895 & -9.99 & 0.63 & 5 & NGC1617\\
F05452-3415 &   0.003070 & 10.50 &  0.713 & 7.22 & 0.69 & 1 & NGC2090\\
F06209-5942 &   0.007560 & 10.89 &  0.516 & 7.40 & 0.53 & 161 & ESO121-G026\\
F07184+8016 &   0.007350 & 11.39 &  0.779 & 5.77 & 0.87 & 2 & NGC2336\\
F07525+6028 &   0.004810 & 10.62 &  0.578 & 8.11 & 0.41 & 3 & NGC2460\\
F08547+0306 &   0.013080 & 11.32 &  0.493 & 6.59 & 0.58 & 1 & NGC2713\\
F09076+0714 &   0.004500 & 11.18 &  0.887 & 6.19 & 0.64 & 3 & NGC2775\\
F09134+7358 &   0.007540 & 11.04 &  0.593 & 5.61 & 0.57 & 3 & IC0529\\
F09186+5111 &   0.002130 & 10.41 &  0.827 & 6.95 & 0.91 & 1 & NGC2841\\
F10068-2849 &   0.003680 & 10.50 &  0.634 & 6.17 & 0.82 & 29 & NGC3137\\
F10312-2711 &   0.011270 & 11.31 &  0.553 & 6.25 & 0.43 & 163 & NGC3285\\
F10346-2718 &   0.009630 & 11.24 &  0.586 & 8.46 & 0.45 & 223 & NGC3312\\
F10456-2034 &   0.013430 & 11.47 &  0.557 & 6.08 & 0.42 & 34 & NGC3450\\
F11089-3636 &   0.008310 & 10.93 &  0.495 & 22.25 & 0.57 & 13 & NGC3573\\
F11163+1321 &   0.002690 & 10.63 &  0.835 & 7.05 & 0.99 & 4 & M065\\
F11549+5339 &   0.003500 & 11.04 &  0.926 & 7.44 & 0.89 & 2 & M109\\
F12015+3210 &   0.002530 & 10.00 &  0.547 & 6.75 & 0.61 & 1 & NGC4062\\
F12133+1325 &   0.000440 &  8.61 &  0.612 & 7.29 & 0.91 & 223 & NGC4216\\
F12149+3805A &  0.000810 &  9.05 &  0.567 & 10.44 & 1.22 & 1 & NGC4244\\
F12194-3531 &   0.009790 & 11.12 &  0.519 & 6.99 & 0.51 & 6 & ESO380-G019\\
F12234+1829 &   0.003070 & 10.04 &  0.483 & 5.41 & 0.57 & 11 & NGC4394\\
F12257+2853 &   0.002210 &  9.87 &  0.541 & 6.30 & 0.60 & 1 & NGC4448\\
F12259+1721 &   0.006520 & 11.06 &  0.666 & 7.10 & 0.72 & 3 & NGC4450\\
F12329+1446 &   0.001620 & 10.30 &  0.890 & 11.02 & 0.74 & 2 & M091\\
F12374-1120 &   0.003420 & 11.76 &  1.296 & 7.34 & 0.95 & 1 & M104\\
F12407+0215 &   0.004450 & 10.69 &  0.647 & 7.47 & 0.49 & 2 & NGC4643\\
F12415-4027 &   0.009850 & 11.19 &  0.552 & 14.12 & 0.56 & 5 & NGC4650\\
F12474-1427 &   0.013060 & 11.41 &  0.539 & 6.31 & 0.36 & 5 & MCG-02-33-017\\
F13191-3622 &   0.001560 & 10.39 &  0.952 & 8.74 & 0.90 & 2 & NGC5102\\
F13331-3312 &   0.013990 & 11.37 &  0.489 & -9.99 & 0.38 & 3 & NGC5220\\
F13447-3041 &   0.014900 & 11.55 &  0.552 & 5.71 & 0.27 & 5 & NGC5292\\
F13550-2904 &   0.008920 & 11.32 &  0.660 & 6.56 & 0.80 & 2 & IC4351\\
F14134+3627 &   0.009590 & 11.12 &  0.528 & 6.96 & 0.79 & 2 & NGC5529\\
F14424+0209 &   0.005750 & 11.36 &  0.870 & 7.45 & 0.89 & 2 & NGC5746\\
F15045+0144 &   0.008530 & 10.99 &  0.514 & 6.26 & 0.64 & 2 & NGC5850\\
F15110-1405 &   0.006640 & 11.03 &  0.643 & 6.09 & 0.59 & 1 & NGC5878\\
F17304+1626 &   0.010400 & 11.24 &  0.553 & 25.21 & 0.49 & 1 & NGC6389\\
F17576-6625 &   0.014480 & 11.56 &  0.569 & -9.99 & 0.45 & 2 & NGC6389\\
F18300-5832 &   0.007460 & 10.84 &  0.497 & 6.94 & 0.76 & 3 & IC4721\\
F19049-6357 &   0.002800 & 11.25 &  1.128 & 9.88 & 1.31 & 2 & NGC6744\\
F19101-6221 &   0.014490 & 11.50 &  0.539 & 14.27 & 0.55 & 3 & IC4831\\
F19139-6035 &   0.012620 & 11.40 &  0.549 & 6.30 & 0.39 & 6 & NGC6769\\
F19227-5503 &   0.010570 & 11.27 &  0.561 & 5.96 & 0.48 & 2 & NGC6788\\
F19588-5613 &   0.014650 & 10.16 &  0.509 & 7.09 & 0.40 & 2 & NGC6848\\
F20134-5257 &   0.009030 & 10.99 &  0.489 & 6.91 & 0.55 & 3 & NGC6887\\
F20210-4348 &   0.009330 & 11.53 &  0.745 & 7.18 & 0.76 & 2 & NGC6902\\
F20346-5217 &   0.015150 & 11.48 &  0.510 & 13.36 & 0.30 & 2 & NGC6935\\
\end{tabular}
\end{table*}

\begin{table*}
\caption{Comparison samples of large galaxies requiring cool dust or normal cirrus components (cont.)}
\begin{tabular}{llllllll}
IRAS name & z & $L_{opt}$ & $log_{10}\theta_{pred}$ & $\theta_{Planck}$ & $log_{10}\theta_{opt}$ & ncirr & name\\
&&&&&&&\\
cool dust galaxies (cont.) &&&&&&&\\
F20597-1700 &   0.004920 & 10.57 &  0.543 & 17.33 & 0.63 & 1 & IC5078\\
F21029-4822 &   0.017200 & 11.63 &  0.529 & 14.32 & 0.43 & 3 & ESO235-G057\\
F21142-6358 &   0.010490 & 11.21 &  0.534 & 6.9 &  0.51 & 4 & IC5096\\
F21596-1909 &   0.008790 & 11.14 &  0.576 & 8.17 & 0.51 & 4 & NGC7183\\
F21598-2103 &   0.008740 & 11.42 &  0.718 & 6.92 & 0.78 & 1 & NGC7184\\
F22061-4724 &   0.005840 & 11.39 &  0.879 & 7.18 & 0.49 & 2 & NGC7213\\
F22369-6644 &   0.010850 & 11.19 &  0.510 & -9.99 & 0.59 & 1 & NGC7329\\
F22450-2234 &   0.011140 & 11.56 &  0.683 & -9.99 & 0.46 & 1 & NGC7377\\
F22521-3955 &   0.005840 & 11.31 &  0.839 & 7.94 & 0.72 & 1 & NGC7410\\
F22543-4339 &   0.005710 & 11.29 &  0.838 & 7.50 & 0.70 & 1 & IC5267\\
F22544-4120 &   0.003130 & 10.48 &  0.694 & 7.42 & 0.98 & 1 & NGC7424\\
F23105-2837 &   0.005220 & 10.60 &  0.532 & 13.64 & 0.52 & 1 & NGC7513\\
F23120-4352 &   0.005320 & 10.80 &  0.624 & -9.99 & 0.66 & 1 & NGC7531\\
&&&&&&&\\
cirrus galaxies &&&&&&&\\
F01319-2940 &   0.004940 & 11.27 &  0.541 & 6.33 & 0.74 & 1 & NGC0613\\
F02207-2127 &   0.005030 & 11.23 &  0.513 & -9.99 & 0.78 & 1 & NGC0908\\
F02441-3029 &   0.004240 & 11.41 &  0.678 & 6.32 & 0.98 & 4 & NGC1097\\
F03075-2045 &   0.005350 & 11.44 &  0.592 & 5.92 & 0.87 & 1 & NGC1232\\
F03207-3723 &   0.005870 & 12.01 &  0.836 & 11.92 & 1.05 & 4 & NGC1316\\
F03316-3618 &   0.005460 & 11.57 &  0.648 & 7.37 & 1.05 & 2 & NGC1365\\
F03404-4722 &   0.003590 & 11.03 &  0.560 & 6.9 &  0.81 & 2 & NGC1433\\
F04022-4329 &   0.002990 & 10.87 &  0.559 & 6.8 &  0.95 & 1 & NGC1512\\
F04101-3300 &   0.003470 & 11.04 &  0.580 & 9.2 &  1.10 & 1 & NGC1532\\
F04188-5503 &   0.005020 & 11.34 &  0.569 & 5.58 & 0.92 & 8 & NGC1566\\
F04449-5920 &   0.004440 & 11.18 &  0.543 & 6.88 & 0.92 & 117 & NGC1672\\
F05035-3802 &   0.004040 & 11.13 &  0.559 & 5.51 & 0.72 & 2 & NGC1792\\
F07365-6924 &   0.004890 & 11.27 &  0.546 & 5.3 &  0.81 & 7 & NGC2442\\
F08491+7824 &   0.004670 & 11.22 &  0.541 & 7.71 & 0.69 & 4 & NGC2655\\
F08495+3336 &   0.001370 & 10.36 &  0.643 & 6.87 & 0.98 & 1 & NGC2683\\
F09293+2143 &   0.001830 & 10.77 &  0.723 & 7.78 & 1.10 & 4 & NGC2903\\
F10126+7338 &   0.009350 & 11.86 &  0.559 & 5.24 & 0.59 & 1 & NGC3147\\
F10407+2511 &   0.001930 & 10.55 &  0.589 & 6.95 & 0.85 & 1 & NGC3344\\
F10413+1157 &   0.002590 & 10.88 &  0.627 & 7.41 & 0.88 & 3 & M095\\
F10441+1205 &   0.002990 & 11.08 &  0.664 & 6.31 & 0.88 & 3 & M096\\
F11032+0014 &   0.002670 & 11.20 &  0.773 & 5.9 &  1.04 & 7 & NGC3521\\
F11176+1315 &   0.002430 & 11.06 &  0.744 & 6.64 & 0.97 & 2 & M066\\
F11176+1351 &   0.002810 & 11.04 &  0.671 & 7.13 & 1.17 & 4 & NGC3628\\
F12006+4448 &   0.002340 & 10.56 &  0.511 & 6.03 & 0.72 & 1 & NGC4051 Seyf1\\
F12163+1441 &   0.008030 & 11.70 &  0.545 & 7.08 & 0.75 & 222 & M099\\
F12193+0445 &   0.005220 & 11.43 &  0.597 & 5.25 & 0.81 & 7 & M061\\
F12203+1605 &   0.005240 & 11.49 &  0.626 & 7.87 & 0.88 & 33 & M100\\
F12239+3130 &   0.002390 & 10.63 &  0.537 & -9.99 & 0.56 & 2 & NGC4414\\
F12294+1441 &   0.007610 & 11.91 &  0.674 & 5.63 & 0.85 & 4 & M088\\
F12334+2814 &   0.002690 & 10.75 &  0.545 & 6.19 & 1.04 & 1 & NGC4559\\
F12410+1151 &   0.004700 & 11.14 &  0.498 & 5.51 & 0.46 & 3 & NGC4647 (or 0.85 M60?)\\
F12464-0823 &   0.004650 & 11.41 &  0.638 & 6.17 & 0.59 & 3 & NGC4699\\
F12498-0055 &   0.004130 & 11.20 &  0.584 & 6.87 & 0.77 & 1 & NGC4753\\
F12542+2157 &   0.001360 & 10.80 &  0.866 & 6.7 &  1.02 & 3 & M064\\
F13086+3719 &   0.003160 & 11.03 &  0.615 & 5.24 & 0.76 & 2 & NGC5005\\
F13170-2708 &   0.007230 & 11.48 &  0.481 & 6.54 & 0.62 & 1 & NGC5078\\
F13277+4727 &   0.001540 & 10.86 &  0.842 & 8.61 & 1.05 & 5 & M051a\\
F13350+0908 &   0.003840 & 10.93 &  0.481 & 6.43 & 0.79 & 1 & NGC5248\\
F22347+3409 &   0.002720 & 11.07 &  0.700 & 6.97 & 1.06 & 53 & NGC7331\\
F23552-3252 &   0.000770 &  9.83 &  0.629 & 6.85 & 0.98 & 1 & NGC7793\\
\end{tabular}
\end{table*}

\section{Results from maps of selected sources}

Mapping data in the submillimetre from the SPIRE instrument (Griffin et al., 2010) on the {\em Herschel Space Observatory} (Pilbratt et al., 2010) is
available for a few of the galaxies identified here as containing cold dust. This allows us to compare the location of the dust
responsible for the submillimetre emission with the location of the starlight.

Two of the sources (NGC5701 and NGC5669) were observed as part of the {\em Herschel} Reference Survey (Boselli et
al. 2010), and a third (NGC3049) was observed by the KINGFISH programme (Kennicutt et al. 2010).
The SPIRE images were obtained from the {\em Herschel} or HerMES data archives.

For each of these objects we used an SDSS $i$ band optical image to trace the optical surface brightness profile from stellar emission after smoothing by a gaussian matched to the size of the 250$\mu$m SPIRE beam, and then compared this to the 250$\mu$m surface brightness profile resulting from dust emission. The surface brightness profiles for this calculation were produced using the the IRAF ellipse package and are plotted in Figs 6L and 7. In all cases the optical and submm surface brightness profiles are clearly different.

This is most obvious in NGC5701, where there is a large ring of cold dust located outside most of the optical light distribution (see Fig. 6R). This ring makes a substantial contribution to the integrated submm emission, but the optical emission from starlight in this region is much less than in the more central regions of this object.

\begin{figure*}
\includegraphics[width=6cm, angle =90]{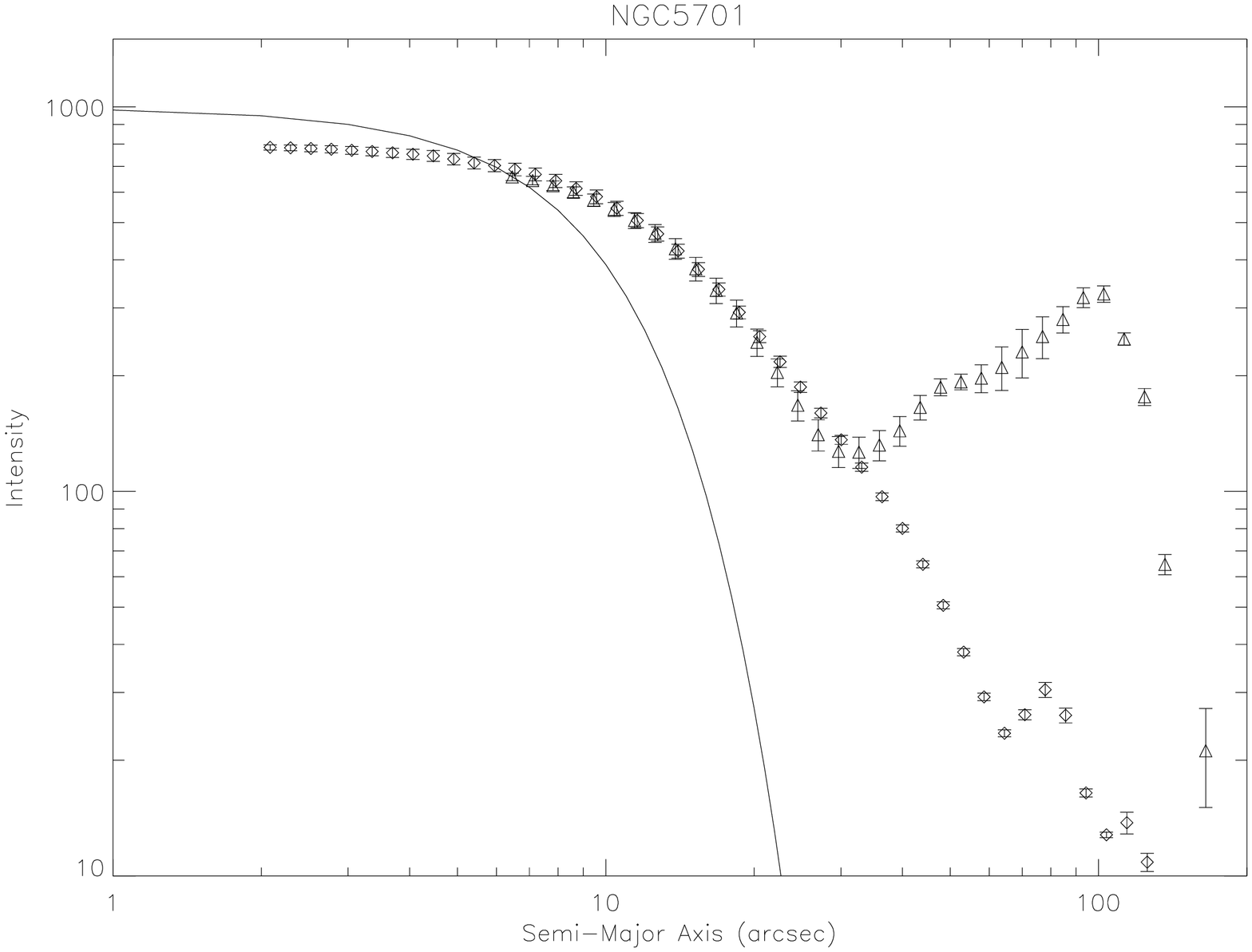}
\caption{
250 $\mu$m and optical intensity profiles of NGC5701. Diamonds indicate the $i$ band surface brightness derived from an SDSS image after it has been smoothed by a gaussian matching the SPIRE beam at 250$\mu$m, triangles show the 250$\mu$m surface brightness from SPIRE observations. The solid line indicates a gaussian model of the SPIRE 250$\mu$m beam as given by the SPIRE Handbook\thanks\url{http://herschel.esac.esa.int/Docs/SPIRE/spire_handbook.pdf}. The true Herschel beam at 250$\mu$m deviates somewhat from a pure Gaussian at radii beyond $\sim$ 20 arcsecondsm at levels $<$ a few percent (Griffen et al. 2013). This is insignificant to the current analysis. The surface brightness scale is arbitrary with the two profiles scaled to the same value at a semi-major axis radius of 10 arcseconds  In this case strong submillimetre
emission arises in a ring of dust lying outside the main stellar distribution, which can be seen in Fig. 6R, leading to a region where the optical-to-submilimetre ratio is low.
}
\end{figure*}

The dust emission in NGC5669 has a larger scale length than the optical emission, with the ratio of submm to optical surface brightness rising all the way to the edge of detectable emission in this object. Examination of the optical image suggests that this dust is associated with extended low surface brightness optical emission away from the galaxy's nucleus and bright spiral arms.

NGC3049 is an edge on spiral, which makes this analysis a little less clear, and we see only a slight hint of changes to the optical to submm surface brightness ratio in the outer parts of this object, or possibly weak submm emission beyond the optically detected emission.

\begin{figure*}
\includegraphics[width=6cm, angle =90]{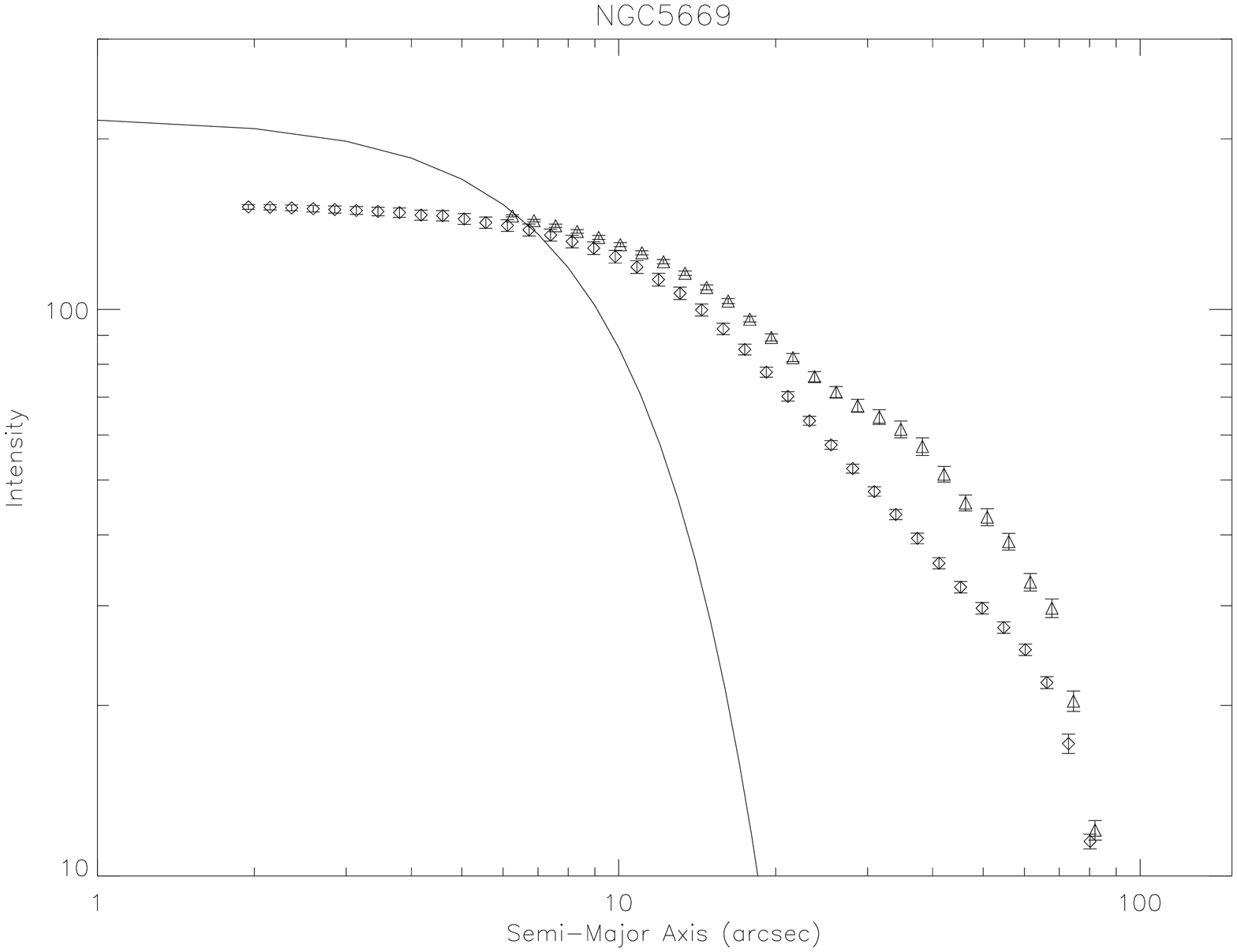}
\includegraphics[width=6cm, angle=90]{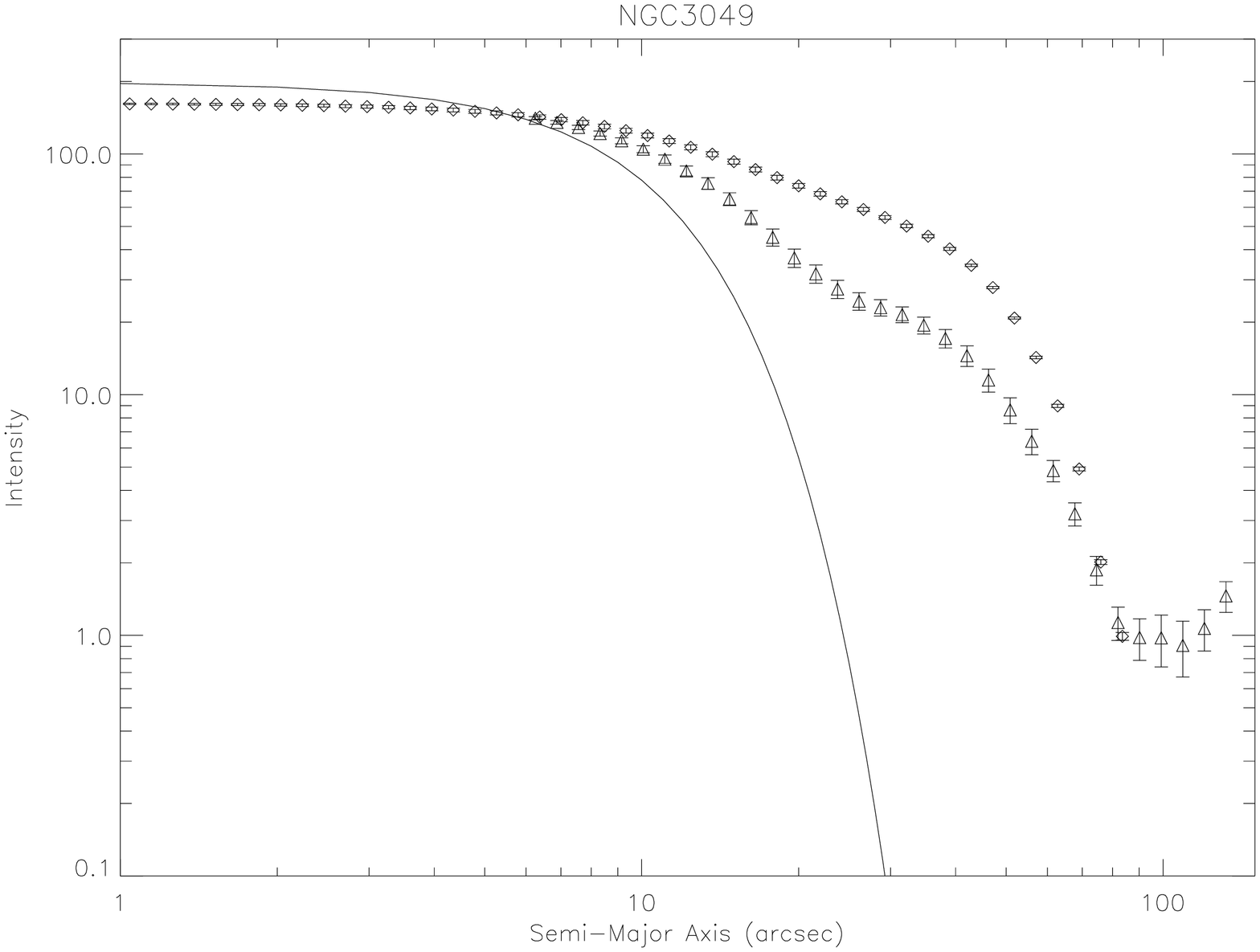}
\caption{
L: 250 $\mu$m and optical intensity profiles of NGC5669. Symbols and scalings as in Fig. 6R.
R: 250 $\mu$m and optical intensity profiles of NGC3049. Symbols and scalings as in Fig. 6R.}
\end{figure*}
 
A further galaxy of interest, NGC1617, classified as cool in Table 2,
was observed as part of the HerMES survey (Oliver et al., 2012) since it lies in the Akari Deep Field South (ADF-S, Matsuura et al 2011).
Figure 9L shows 250 $\mu$m contours superposed on an r-band image (from Hameed and Devereux 1999), while Fig 9R shows 250 $\mu$m and r-band
intensity profiles.  In this case a significant fraction of the submillimetre emission arises from
two dust blobs that lie along the major axis to the north-west.

\begin{figure*}
\includegraphics[width=6cm, angle =0]{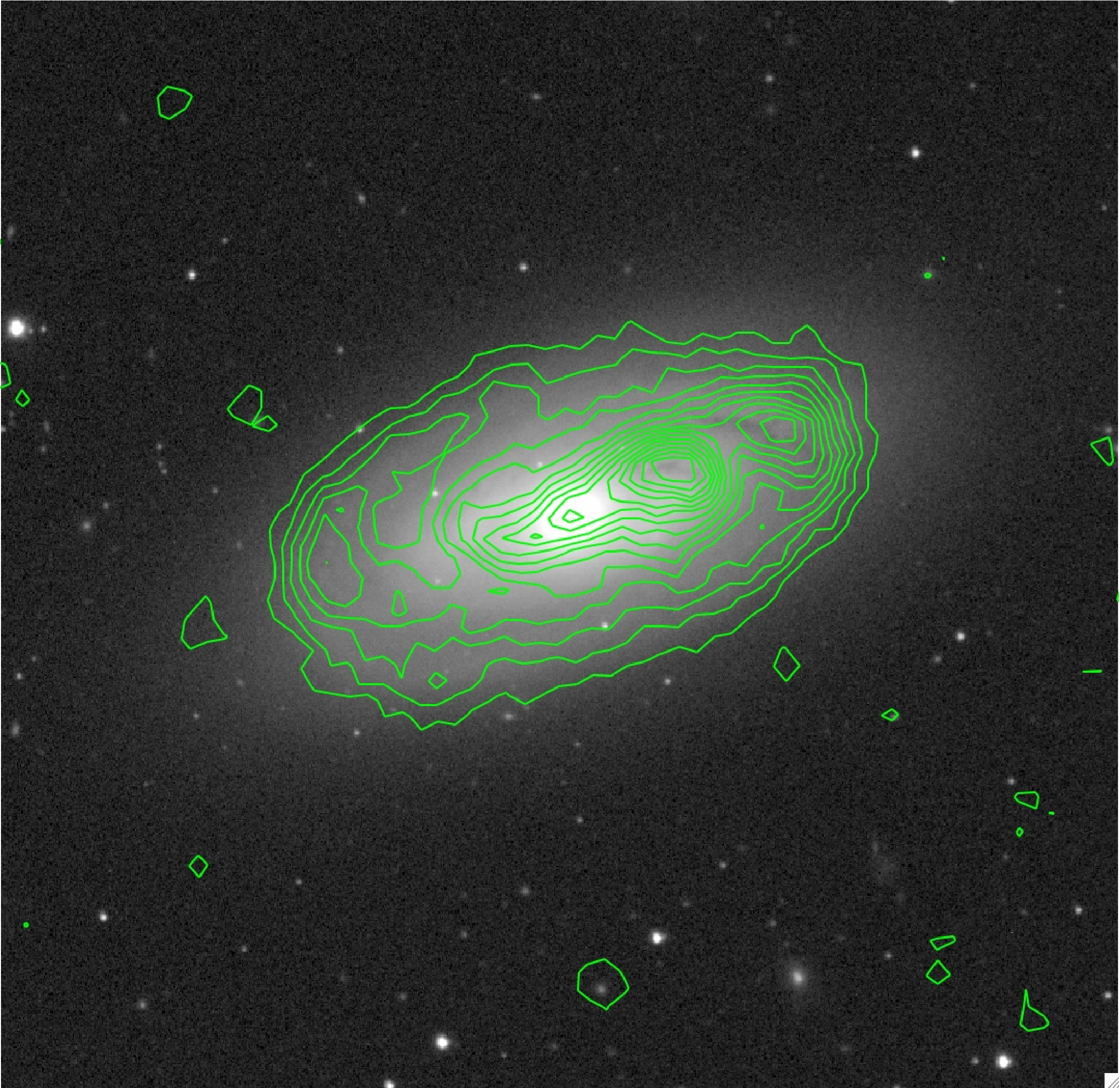}
\includegraphics[width=6cm, angle=90]{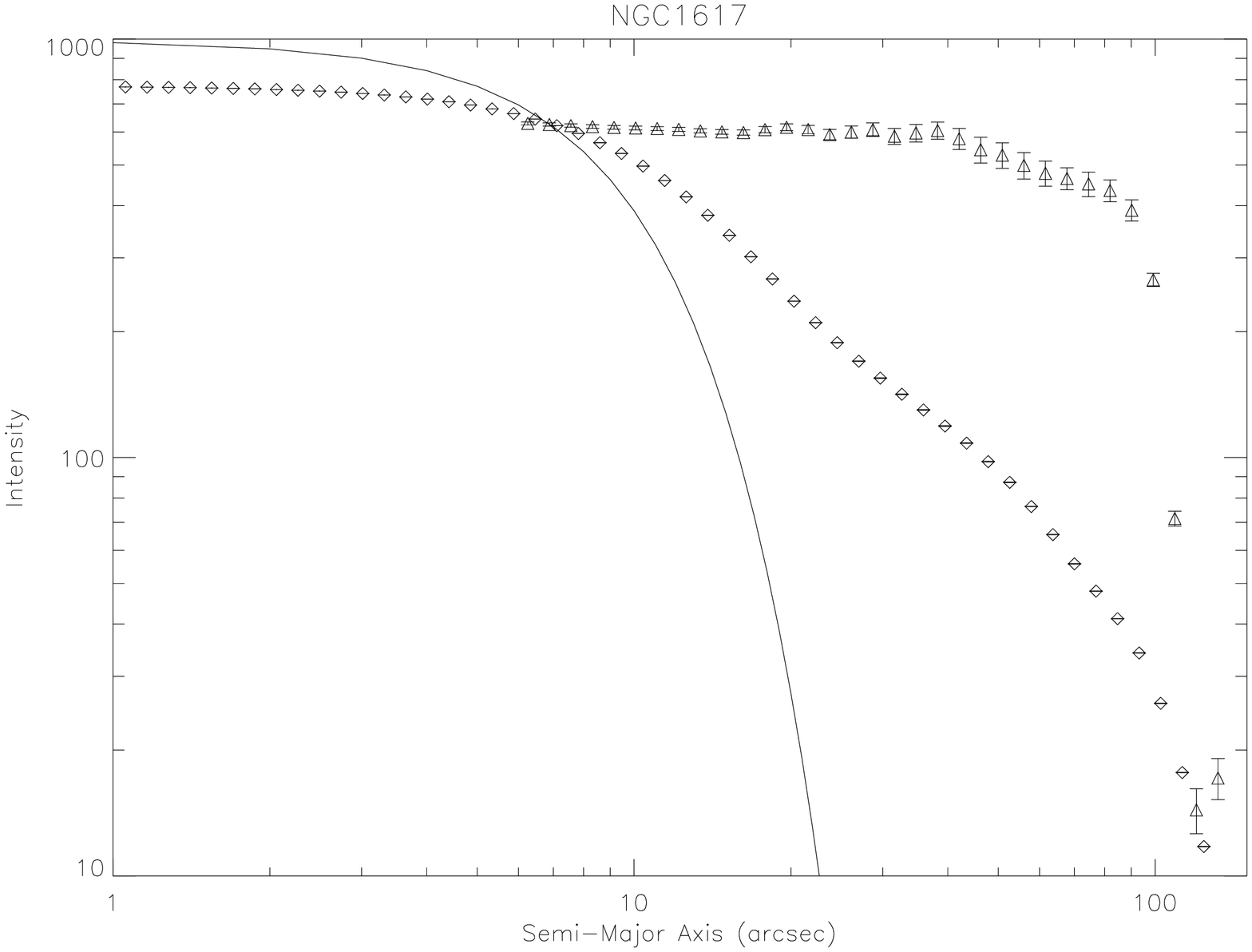}
\caption{
L: 250 $\mu$m contours of NGC1617 superposed on r-band image.
R: 250 $\mu$m and optical intensity profiles of NGC1617. Symbols and scalings as in Fig. 6R.}
\end{figure*}
 
\section{Discussion and conclusions}

The analysis of {\it Planck} 350 $\mu$m data shows very clearly that the galaxies whose SEDs require a cold dust component
are indeed larger than normal galaxies and by an amount consistent with the interpretation that the surface
brightness of the illuminating starlight is lower.
The fact that this phenomenon of galaxies with cold dust did not show up in follow-up of IRAS
galaxies refelects the fact that 60 $\mu$m selection biases samples towards galaxies with warmer dust.
It needed the submillimetre sensitivity of {\it Herschel} and {\it Planck} to uncover this phenomenon.

The limited submillimetre mapping data available for these cold dust galaxies shows that a variety of
geometries may be involved.  NGC5701 shows a large ring of cold dust located outside most of the optical
light distribution but in NGC3049 the cold dust appears to be located closer to the centre of the galaxy.

It is unlucky that none of our cold dust galaxies is in the sample of nearby galaxies for which Bendo et al
(2015) have studied the 160/250 and 250/350 $\mu$m surface brightness profiles in detail.  The next step
will be to map some of these cold dust galaxies with ground-based submillimetre telescopes.

%

\section{Acknowledgements}

{\it Herschel} is an ESA space observatory with science instruments provided by European-led
Principal Investigator consortia and with important participation from NASA. SPIRE has
been developed by a consortium of institutes led by Cardiff University (UK) and
including Univ. Lethbridge (Canada); NAOC (China); CEA, LAM (France); IFSI, Univ.
Padua (Italy); IAC (Spain); Stockholm Observatory (Sweden); Imperial College London,
RAL, UCL-MSSL, UKATC, Univ. Sussex (UK); and Caltech, JPL, NHSC, Univ. Colorado (USA).
This development has been supported by national funding agencies: CSA (Canada); NAOC
(China); CEA, CNES, CNRS (France); ASI (Italy); MCINN (Spain); SNSB (Sweden); STFC,
UKSA (UK); and NASA (USA).

Based in part on observations obtained with {\em Planck}
([LINK]http://www.esa.int/{\em Planck}), an ESA science mission with instruments and
contributions directly funded by ESA Member States, NASA, and Canada.

We thank George Bendo and an anonymous referee for helpful comments.



\begin{thebibliography}{99}

\bibitem{} Ade P.A.R. et al, 2011, Planck early results XVI, AA 536, 16

\bibitem{} Bendo G.J. et al, 2015, MNRAS 448, 135

\bibitem{} Boselli A. et al, 2010, PASP 122, 261

\bibitem{} Bourne N. et al, 2013, MNRAS 436, 479

\bibitem{} Efstathiou A., Rowan-Robinson M., 2003, MNRAS 343, 322

\bibitem{} Efstathiou A., Rowan-Robinson M., Siebenmorgen R., 2000, MNRAS 313, 734

\bibitem{} Galametz M. et al, 2012, MNRAS 425, 763

\bibitem{} Griffin M.J. et al, 2010, AA 518, L3

\bibitem{} Griffin M.J. et al, 2013, MNRAS 434, 992

\bibitem{} Hameed S., Devereux N., 1999, AJ 118, 730

\bibitem{} Ibar E. et al, 2013, MNRAS 449, 2498

\bibitem{} Kennicutt R.C. et al, 2010, PASP 123, 1347



\bibitem{} Pilbratt G. et al, 2010, AA 518, L1

\bibitem{} Rice W., Lonsdale C.J., Soifer B.T., Neugebauer G., Kopan E.L., Lloyd L.A., de Jong T., 
Habing H.J., 1988, ApJS 68, 91

\bibitem{} Rowan-Robinson M., 1992, MNRAS 258, 787

\bibitem{} Rowan-Robinson M. et al, 2005, AJ 129, 1183

\bibitem{} Rowan-Robinson M. et al, 2008, MNRAS 386, 697

\bibitem{} Rowan-Robinson M. et al, 2010, MNRAS 409, 2

\bibitem{} Rowan-Robinson M. et al, 2013, MNRAS 428, 1958

\bibitem{} Rowan-Robinson M. et al, 2014, MNRAS 445, 3848

\bibitem{} Smith D.J.B. et al, 2012, MNRAS 427, 703

\bibitem{} Symeonidis M. et al, 2013, MNRAS 431, 2317

\bibitem{} Wang L., Rowan-Robinson M., 2009, MNRAS 398, 109

\bibitem{} Wang L., Rowan-Robinson M., Norberg P., Heinis S., Han J., 2014, MNRAS 442, 2739


\end{thebibliography}
\end{document}